\documentclass[aps,amsmath,amssymb,notitlepage,nofootinbib,twocolumn]{revtex4-1}
\usepackage{slashed}
\usepackage{graphicx}
\usepackage[pdftex,final]{hyperref}

\newcommand{\be}{\begin{equation}}
\newcommand{\ee}{\end{equation}}
\def\beq{\begin{equation}}
\def\eeq{\end{equation}}
\def\bea{\arraycolsep .1em \begin{eqnarray}}
\def\eea{\end{eqnarray}}
\def\s0#1#2{\mbox{\small{$ \frac{#1}{#2} $}}}
\def\0#1#2{\frac{#1}{#2}}
\def\eq#1{(\ref{#1})}
\def\step{\\[-1.5ex]}

\renewcommand{\d}{\mathrm{d}}
\newcommand{\psib}{\bar{\psi}}
\newcommand{\la}[1]{\lambda_{#1}}

\begin{document}

\title{Line of  Fixed Points  in Gross-Neveu Theories}

\author{Charlie Cresswell-Hogg}
\email{c.cresswell-hogg@sussex.ac.uk}
\author{Daniel F.~Litim}
\email{d.litim@sussex.ac.uk}
\affiliation{Department of Physics and Astronomy, University of Sussex, Brighton, BN1 9QH, U.K.}

\begin{abstract}
In the limit of many fermion flavors it is demonstrated that the sextic Gross-Neveu theory  in three dimensions displays a line of interacting UV fixed points,  characterised by an exactly marginal sextic interaction. We  determine the conformal window of UV-complete theories, universal scaling dimensions, and the phase diagram using  renormalisation group methods. Massless theories arise naturally, and  the    generation of mass proceeds without the breaking of a discrete symmetry.
Striking similarities with critical scalar theories  at large~$N$ are highlighted, and  implications from the viewpoint of conformal field theory and the AdS/CFT conjecture are indicated.
\end{abstract}

\maketitle

{\it Introduction.---} 
The large $N$ limit, where $N$ denotes the number of particle species or flavors,  is an important  tool
in quantum   and statistical physics \cite{Moshe:2003xn}. 
Large $N$ limits  often enable rigorous  control over  fluctuations and critical points \cite{Veneziano:1979ec,Banks:1981nn,Litim:2014uca}
including  beyond perturbation theory \cite{David:1984we,Bond:2022xvr}, and offer wide-ranging applications  from proofs of non-perturbative renormalisability \cite{deCalan:1991km} and new quantum critical points \cite{Bond:2017lnq,Bond:2017suy} or symmetry breaking mechanisms \cite{Bardeen:1983rv}, to  equivalences  \cite{Lovelace:1982hz,Bond:2019npq}, dualities \cite{Aharony:2012ns,Seiberg:2016gmd}, and  the AdS/CFT conjecture  \cite{Maldacena:1997re}, which further relates large $N$  field theories  to   higher-dimensional duals \cite{Giombi:2012ms}.

The $3d$ Gross-Neveu theory of interacting fermions, originally introduced
for the study of chiral symmetry breaking   \cite{Gross:1974jv}, is an important toy model in particle and condensed matter physics.  
Its  four-fermion coupling $\sim ( \psib_a \psi_a )^2$ 
 makes the theory   perturbatively non-renormalisable, yet
non-perturbatively, it develops an interacting  fixed point
\cite{Rosenstein:1988pt} including at large $N$ \cite{deCalan:1991km}, which renders the theory UV complete and predictive at all scales~\cite{Braun:2011pp,Jakovac:2014lqa}.  Besides  offering a benchmark for  3d conformal field theories and chiral symmetry breaking,
it has also  been shown to be dual 
to  Vassiliev's higher spin theories under the AdS${}_4$/CFT${}_3$ conjecture \cite{Sezgin:2003pt,Giombi:2012ms},
and serves as a toy model for  asymptotic safety of gravity \cite{Braun:2010tt}.

In this Letter, we investigate  large $N$ Gross-Neveu theories in the absence of   chiral symmetry.\footnote{In a slight abuse of language, we refer to   chiral symmetry and parity interchangeably.} 
 The generation of mass is no longer protected by symmetry, and the impact of chirally odd interactions  such as $\sim (\psib_a \psi_a )^3$ 
 on the short distance physics
 needs to be clarified. 
Our method of choice is functional renormalisation \cite{Polchinski:1983gv,Wetterich:1992yh,Ellwanger:1993mw,Morris:1993qb}, which, in combination with the large $N$ limit, enables a non-perturbative study of   fixed points, scaling dimensions, and UV-IR connecting trajectories \cite{Tetradis:1995br,DAttanasio:1997yph,Litim:2011bf,Litim:2017cnl,Litim:2018pxe} in a purely fermionic formulation \cite{Jakovac:2013jua}.
 We thereby determine   all UV complete Gross-Neveu theories, with or without chiral symmetry.
 We also  uncover striking similarities between large $N$ fermionic and  large $N$ bosonic theories,  and indicate  links with conformal field theory and higher spin gauge theories.

 {\it Sextic Gross-Neveu Theory.---}
We consider $U(N)$ symmetric classical actions for $N$ Dirac fermions $\psi_a$ in three Euclidean dimensions, given by
\be\label{eq:classical_action}
S_{\rm f} = \! \!\int_x  
\Big\{ \psib_a( \slashed \partial +M)\psi_a + \frac G2 ( \psib_a \psi_a )^2 + \frac{H}{3!}  ( \psib_a \psi_a )^3 \Big\}.
\ee
In addition to the kinetic term and a mass term $M$, we observe a  four-fermion interaction with coupling $G$, and a six-fermion interaction with coupling $H$. 
In the limit  $M=H=0$  the theory reduces to the standard Gross-Neveu model~\cite{Gross:1974jv} with manifest  invariance under 
\be\label{eq:discrete_symmetry}
\psi \to \gamma^5 \psi, \quad \psib \to - \psib \gamma^5\,,\quad \psib \psi\to -\psib \psi\,,
\ee
which corresponds to chiral symmetry (parity)  in even (odd) dimensions. In the chiral limit, interactions include
even powers of the bilinear $( \psib_a \psi_a )$, the leading one being the  four-fermion (4F) interaction with coupling $G$. Owing to its negative canonical mass dimension, $[G]=-1$, the theory is perturbatively non-renormalisable by power counting,  yet  non-perturbatively renormalisable due to the existence of an interacting ultraviolet  fixed point \cite{Rosenstein:1988pt,deCalan:1991km}. The 4F coupling becomes a relevant interaction while all higher order chirally-invariant interactions remain irrelevant. The discrete symmetry \eq{eq:discrete_symmetry} further entails that the theory is fundamentally massless, although mass can be generated dynamically in the infrared. 

Without  chiral symmetry, 
mass terms $M\neq 0$ and
new interactions such as odd powers in $( \psib_a \psi_a )$ become available, the leading one being the six-fermion (6F) coupling $H$. 
It is the central purpose of this work to establish that theories with $H\neq 0$ can be defined fundamentally, despite of the fact 
that negative  mass dimension of the 6F coupling $[H]=-3$ implies that the theory is power-counting  non-renormalisable in perturbation theory. 
\step

{\it Renormalisation Group.---}
To achieve our claims, we investigate 
the theory~\eqref{eq:classical_action} non-perturbatively with the help of  Wilson's 
renormalisation group~\cite{Polchinski:1983gv,Wetterich:1992yh,Ellwanger:1993mw,Morris:1993qb} based on the successive integrating-out of momentum modes from a path integral representation of the   theory.  The scale-dependence of the quantum effective action $\Gamma_k$  is given by  an exact identity \cite{Wetterich:1992yh}
\be
\label{eq:wetterich}
\partial_t \Gamma_k = \tfrac{1}{2} {\rm STr} \left\{ \big[ \Gamma_k^{(2)} + R_k \big]^{-1} \cdot \partial_t R_k \right\}\,,
\ee
where $k$ denotes the RG momentum scale with $t=\ln k$, $\Gamma_k^{(2)}$ denotes the matrix of second functional derivatives, and the supertrace STr stands for a sum over all momenta and fields, also accounting for relative minus signs through fermionic degrees of freedom. 
As a function of $k$, the flow \eq{eq:wetterich} interpolates between the microscopic action in the ultraviolet $(1/k\to 0)$ and the full quantum effective action in the infrared $(k\to 0)$. Within a few constraints, the infrared cutoff function $R_k(q)$ can chosen freely~\cite{Litim:2000ci,Litim:2001up,Litim:2001fd}. 
It ensures that the propagation of momentum modes is suppressed for $q^2\ll k^2$ and that the flow remains finite and well-defined for all scales.
Further, Wetterich's RG  \eq{eq:wetterich} reproduces standard perturbation theory at weak coupling, and 
is related to Polchinski's  RG (based on an UV cutoff) \cite{Polchinski:1983gv}  by a duality transform \cite{DAttanasio:1997yph,Litim:2018pxe}. 
We use \eq{eq:wetterich} to identify interacting UV fixed points in the fermionic theory \eq{eq:classical_action}. For similar studies in scalar and supersymmetric theories, see  \cite{Litim:2011bf,Heilmann:2012yf,Jakovac:2014lqa,Litim:2016hlb,Litim:2017cnl,Litim:2018pxe}.
\step

{\it Fermionic Flow.---} 
The theory \eq{eq:classical_action} is investigated 
by writing its effective action as
\be
\label{eq:eff_action}
\Gamma_k [ \psib, \psi ] = \int d^3x \Big\{ \psib_a \slashed{\partial} \psi_a + V_k \big( \psib_a \psi_a \big) \Big\}\,,
\ee
where the function $V_k$ accounts for all non-derivative interactions. 
In the large-$N$ limit, the functional flow \eq{eq:wetterich} does not generate derivative
interactions other than those already present in the initial action~\cite{DAttanasio:1997yph}.  
Consequently, the anomalous dimension vanishes, and  \eq{eq:eff_action} is exact under the flow \eq{eq:wetterich}, and valid for all scales.
The  flow  for the function $V_k$ is obtained by inserting the ansatz~\eqref{eq:eff_action} into~\eqref{eq:wetterich} and projecting onto constant fields \cite{Jakovac:2013jua,BMB}.
Introducing dimensionless variables 
$z = \psib_a \psi_a/k^2$ and
$v ( z; t ) = V_k ( \psib_a \psi_a )/k^3$ gives
\be
\label{eq:large-N_flow}
\partial_t v +3 v - 2 z v' =-3 A \int_0^\infty\!\! \d y\,  \frac{ y^{5/2}(1+ r) \, r' (y)}{y(1+r)^2 + (v')^2} \,,
\ee
where $y=q^2/k^2$, $v'=\partial_z v$, and $R_k(q)=\slashed q\, r(y)$ with $r(y)$ the cutoff shape function \cite{Litim:2000ci,Litim:2001up}.
The  terms on the left hand side 
account for the canonical scaling of $v$ and $z$, while the integral on the right hand side, a remnant of the operator trace in \eq{eq:wetterich}, is due to quantum fluctuations. 
We remove the constant $A= 4\,N\,S_3 /  ( 2 \pi )^3 $ with $S_3$ the surface area of a unit $3$-sphere by rescaling it into $v\to v/A$ and $z\to z/A$ implying that couplings are  now   measured in units of perturbative loop factors, in accord with na\"ive dimensional analysis  \cite{Weinberg:1978kz}.
\step

{\it Running Couplings and Line of Fixed Points.---} Fixed points are the $t$-independent solutions $v_*(z)$ of $\partial_t v=0$.
The free theory of non-interacting fermions  corresponds to $v'_*=0$.  
To find interacting fixed points, we expand
$v ( z ) = \sum_{n = 1}^\infty \la{n} z^n/{n!}$ in terms of  couplings $\la{n}$ describing the $2n$ fermion (2nF) interactions,
with $\la{1,2,3}$ the dimensionless counterparts of $M,G,H$ in \eq{eq:classical_action}.
Their flows $\partial_t\la{n}$  follow from \eq{eq:large-N_flow} by projection.
For concreteness, we use an optimised regulator $r ( y ) = ({1}/{\sqrt{y}} - 1 )  \theta \left( 1 - y \right)$ which permits simple analytical expressions~\cite{Litim:2000ci,Litim:2001up,Litim:2001fd,Litim:2002cf}; key results are independent of this choice. 

Starting with the dimensionless mass, we find that its flow 
is driven by the mass  and the 4F coupling,
\beq\label{beta1}
\partial_t\la{1} = - \la{1} \left(1- \frac{2  \la{2}}{( 1 + \la{1}^2 )^2}\right)\,.
\eeq
We observe that the fermion mass  is natural \cite{tHooft:1979rat} in that quantum fluctuations or interactions cannot switch it on
if it has been set to zero initially, even if chiral symmetry is absent. Contributions  proportional to the chirally-odd 6F coupling do not arise because the exact flow \eq{eq:large-N_flow} does not involve $v''$ on its right-hand side. Turning to the 4F coupling, we observe that its flow is driven by the mass, and the 4F and  6F couplings,
\beq\label{beta2}
\partial_t\la{2}=  \la{2} + \frac{( 2 - 6 \la{1}^2 ) \la{2}^2}{( 1 + \la{1}^2 )^3}+ \frac{2 \la{1} \la{3}}{( 1 + \la{1}^2 )^2} \,.
\eeq
Fluctuations and interactions can switch on 4F interactions even if they were set to vanish initially. 
However, any dependence on the  chirally odd interactions drop out as soon as the mass term vanishes, leading to
\beq\label{beta2projected}
\partial_t\la{2} \big|_{\la{1}^*=0}=  \la{2}\left(1 + {2  \la{2}}\right)\,.
\eeq
Besides the free fixed point   
we  observe an interacting one for the 4F coupling $\la{2}^*=-\s012\,.$ It coincides with  the well-known ultraviolet fixed point in the chirally symmetric 
limit
~\cite{Wilson:1972cf,Parisi:1975im,Gawedzki:1985ed,Gawedzki:1985jn,Rosenstein:1988pt,deCalan:1991km,Braun:2010tt,Jakovac:2014lqa}.

We now turn to the 6F coupling, which is the lowest-order 
chirally odd interaction. Its RG flow is driven by the mass, and the 4F, 6F and 8F couplings,
\beq\label{beta3}
\partial_t\la{3} = 3 \la{3} + 6 \la{2}  \la{3}\frac{ 1 - 3 \la{1}^2 }{( 1 + \la{1}^2 )^3}+ 2 \la{1}\frac{ \la{4}-12 \frac{ \la{2}^3(1-\la{1}^2 ) }{( 1 + \la{1}^2 )^2}}{( 1 + \la{1}^2 )^2} .
\eeq
Fluctuations  can switch on 6F interactions even if they were  absent initially. 
However, provided the mass term vanishes,  
the beta function simplifies into 
\beq\label{beta3projected}
\partial_t\la{3}  |_{\la{1}^*=0} = 3( 1 + 2 \la{2} ) \la{3}\,.
\eeq
Most notably, $\partial_t\la{3}$  
vanishes identically, and independently of $\la{3}$, provided that the mass and the 4F coupling take their respective  fixed point values. 
In other words, quantum fluctuations 
have turned the perturbatively irrelevant 6F coupling into an exactly marginal one,
which takes the role of a new fundamental parameter. 

For all higher order interactions, solving $\partial_t\la{n\ge 4}=0$
gives the $2n$F couplings at a fixed point as functions of the 6F coupling, 
$\la{n\ge 4}^* = P_n ( \la{3}^* )$, with $P_n$  polynomials of degree $n - 2$ with vanishing constant term  for  $n$ odd.
Then, all 
$2n$F interactions with $n>3$ are  found to be  irrelevant  non-perturbatively
without offering new fundamentally
 free parameters. 
We conclude that the theory displays a line of  critical points parametrized by $\la{3}^*$, which reduces to 
known results in the chiral limit.
\step

{\it Global Fixed Points.---}  
Next, we show that the line of interacting fixed points is limited to a finite range in $\la{3}^*$ values.
To that end,  we integrate \eq{eq:large-N_flow} at a fixed point directly, without first resorting to a polynomial expansion.
Using the method of characteristics~\cite{Tetradis:1995br,Heilmann:2012yf,Litim:2017cnl,Litim:2018pxe} leads to solutions of the form $z=z(v')$ with
\begin{align}\label{solution}
z =   4\,\la{3}^*\, (v')^2 -v' \left[ \0{{}\ 2+3(v')^2}{1 + (v')^2} + 3 v' \arctan v' \right]\,,
\end{align}
whose validity is confirmed by direct inspection. Further, inverting the solution into $v'(z)$ and expanding it in a power series for small $z$, we recover all $2nF$ fixed point couplings determined previously. The virtue of the closed expression \eq{solution} is its validity even beyond the radius of convergence of polynomial expansions. 

\begin{figure}[t]
\includegraphics[width=.8\linewidth]{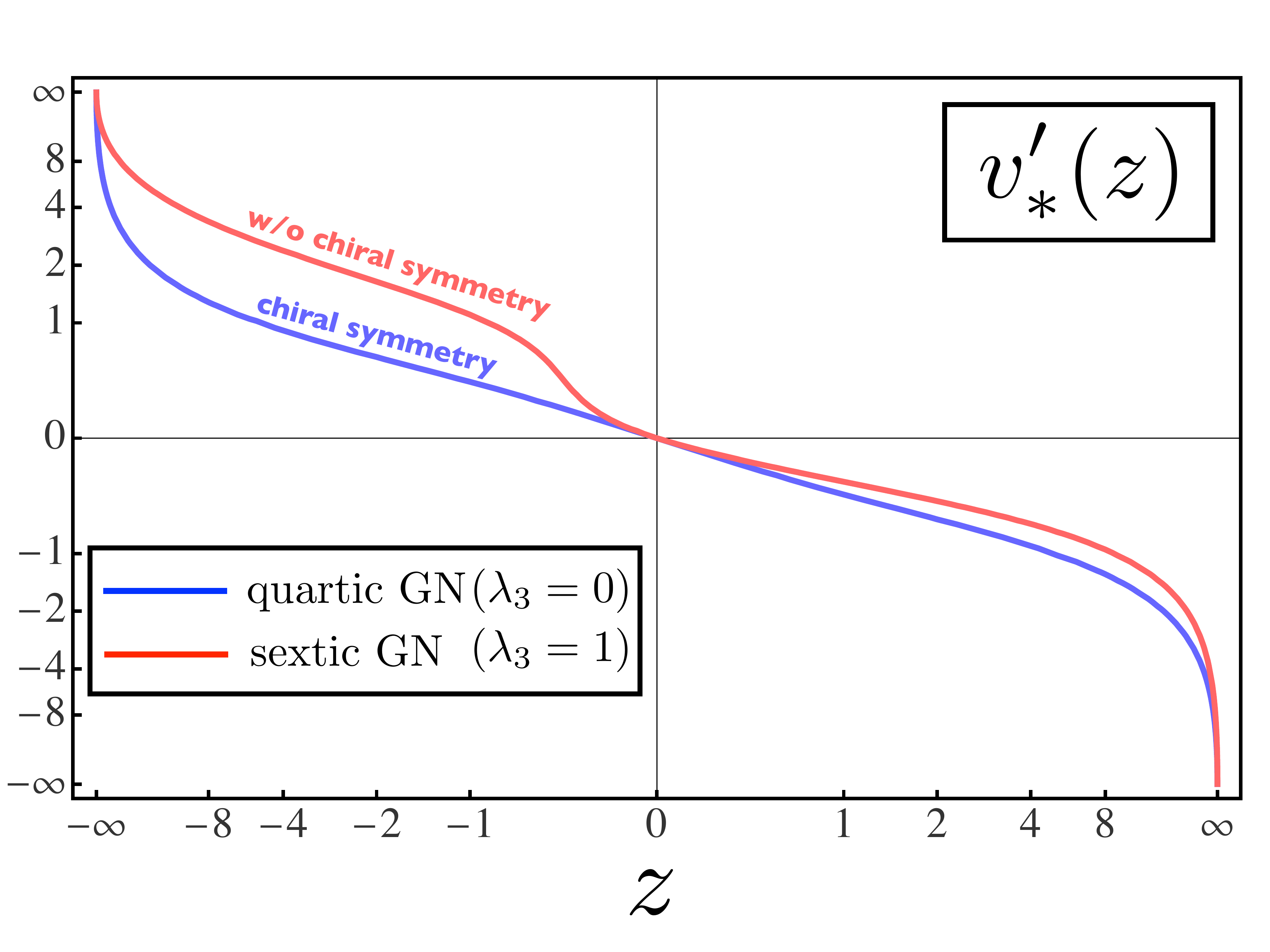}
\caption{Shown are ultraviolet fixed point solutions  $v'_*(z)$ for all fields, comparing  UV-complete quartic and sextic  Gross-Neveu theories (axes scaled  as $x\to \frac{x}{2+|x|}$).}
\label{global}
\end{figure}

As a physical requirement, we  demand  fixed points with \eq{solution} to exist  ``globally'', meaning for all fields $z$.
 Interestingly, this requirement is not empty: if and only if  $|\la{3}^*|$ remains below a critical vale $\la{3}^{\rm crit}$, the fixed point $v(z)$ exists for all values of the field $z$ \cite{Cresswell-Hogg:2022lez} (similar  restrictions are known from other  large-$N$ critical theories at strong coupling \cite{Litim:2011bf,Heilmann:2012yf,Jakovac:2014lqa,Litim:2016hlb,Litim:2017cnl,Litim:2018pxe}). This  can also be appreciated from recursively solving all $\partial_t\la{n\ge 1}=0$ to find $\la{n+1}^*$ as functions of $\la{1}^*$. Resumming the expressions shows that  global solutions do not exist 
if  $\la{1}^*\neq 0$, leading to the same constraint.  
All in all, we  conclude that 
\beq\label{line}
\la{1}^*=0\,,\quad
\la{2}^*=-\s012\,,\quad
0\le |\la{3}^*|< \lambda_3^{\rm crit}
\eeq
characterises the complete set of interacting and globally well-defined fixed points. 
While the values $\lambda_2^*$ and $\lambda_3^{\rm crit}$ are non-universal ($\la{3}^{\rm crit}= \s0{3\pi}{8}$ here), we have checked by varying the cutoff shape function in \eqref{eq:large-N_flow} that the existence of a line with an upper bound is universal~\cite{Litim:2017cnl,Cresswell-Hogg:2022lez}.
Chiral symmetry is  only available if $\la{3}^* = 0$.
Fig.~\ref{global} illustrates two examples for viable fixed points.
\step

\begin{figure}[t]
\includegraphics[width=.915\linewidth]{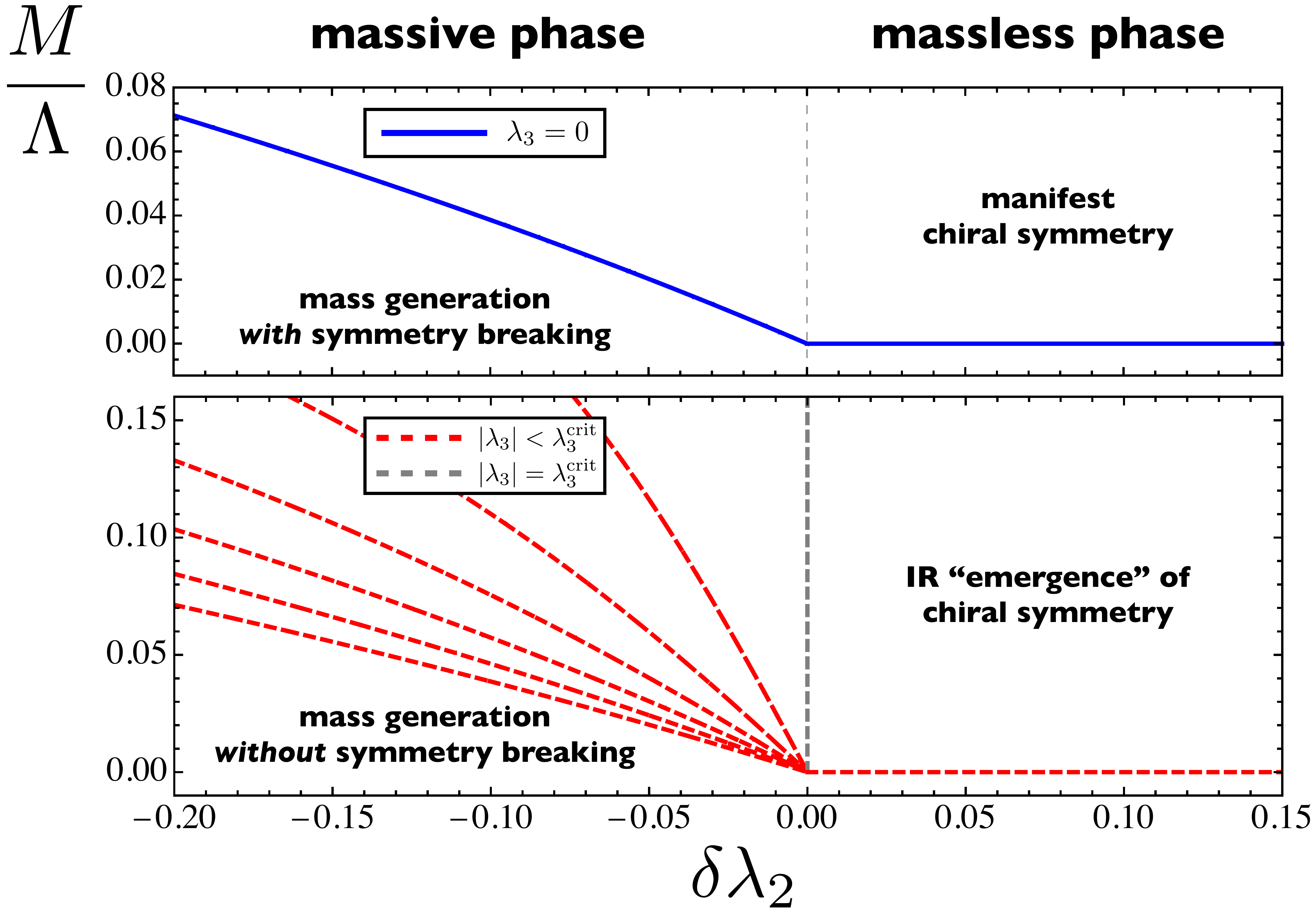}
\vskip-.35cm
\caption{The dynamical generation of a fermion mass $M$ 
takes the form of a second order quantum phase transition in $\delta\la{2}(\Lambda)$, irrespective of whether 
the underlying Lagrangian is chirally invariant (upper panel) or not (lower panel).
}
\label{mass}
\end{figure}

{\it Scaling Dimensions   and Fundamental Parameters.---} In the vicinity of a fixed point, \eq{eq:eff_action} and  \eq{eq:large-N_flow} can be expanded in a basis of  scaling operators $\mathcal{O}_n$ which scale   $\sim k^{-3+\vartheta_n}$ with universal exponents $\vartheta_n$.
In the free theory which is an infrared fixed point, we have $\mathcal{O}_n \sim (\psib\psi)^n$ and    exponents 
are given by (minus) their canonical mass dimensions,
\beq\label{IR}
\vartheta^{\rm (IR)}_n=2n-3\quad\quad (n\ge 1)\,,
\eeq
with the mass being the sole relevant perturbation $\vartheta^{\rm (IR)}_1$.  
In turn, the  fixed points \eq{line} are all UV.
For the universal scaling dimensions of small perturbations we find
\beq\label{UV}
\vartheta^{\rm (UV)}_n=n-3\quad\quad (n\ge 1)\,.
\eeq
The relevant perturbations $\vartheta^{\rm (UV)}_{1,2}$ and  the marginal $\vartheta^{\rm (UV)}_{3}$ relate  to the mass, the 4F, and  the 6F interaction, respectively. Fluctuations have shifted scaling dimensions  substantially away from canonical values \eq{IR},
\beq
\frac{\vartheta^{\rm (IR)}_n-\vartheta^{\rm (UV)}_n}{\vartheta^{\rm (IR)}_n}=
\frac{n+1}{2n-1}\,,\eeq
a testament to the theory being strongly coupled in the UV. 
We note that even though scaling dimensions are the same along the line of fixed points, the underlying discrete symmetries are not.
Further, at the endpoints of the critical line ($|\la{3}^*|= \lambda_3^{\rm crit}$) scale symmetry is broken spontaneously, akin of the Bardeen-Moshe-Bander phenomenon \cite{Bardeen:1983rv,Bardeen:1983st,David:1984we}, and scaling dimensions deviate from the values given in  \eq{UV} (see \cite{Cresswell-Hogg:2022lez} for a detailed analysis).  
We conclude that  the perturbatively non-renormalisable theories \eq{eq:classical_action}  are well-defined due to the line of fixed points  \eq{line}, and  fundamentally characterised  by the exactly marginal coupling  $\la{3}^*$   and  the relevant perturbations $\delta\la{1}=\la{1}-\la{1}^*$ and $\delta\la{2}=\la{2}-\la{2}^*$ at the high scale $\Lambda$.

{\it Generation of Mass.---}  
The absence of chiral symmetry on the level of the fundamental action implies that mass can be generated explicitly as soon as $\delta\la{1}\neq 0$ at the high scale. If so, this relevant perturbation triggers an RG flow towards the IR, for all couplings, which invariably entails a massive fermionic theory.   
  
\begin{figure}[t]
\includegraphics[width=.8\linewidth]{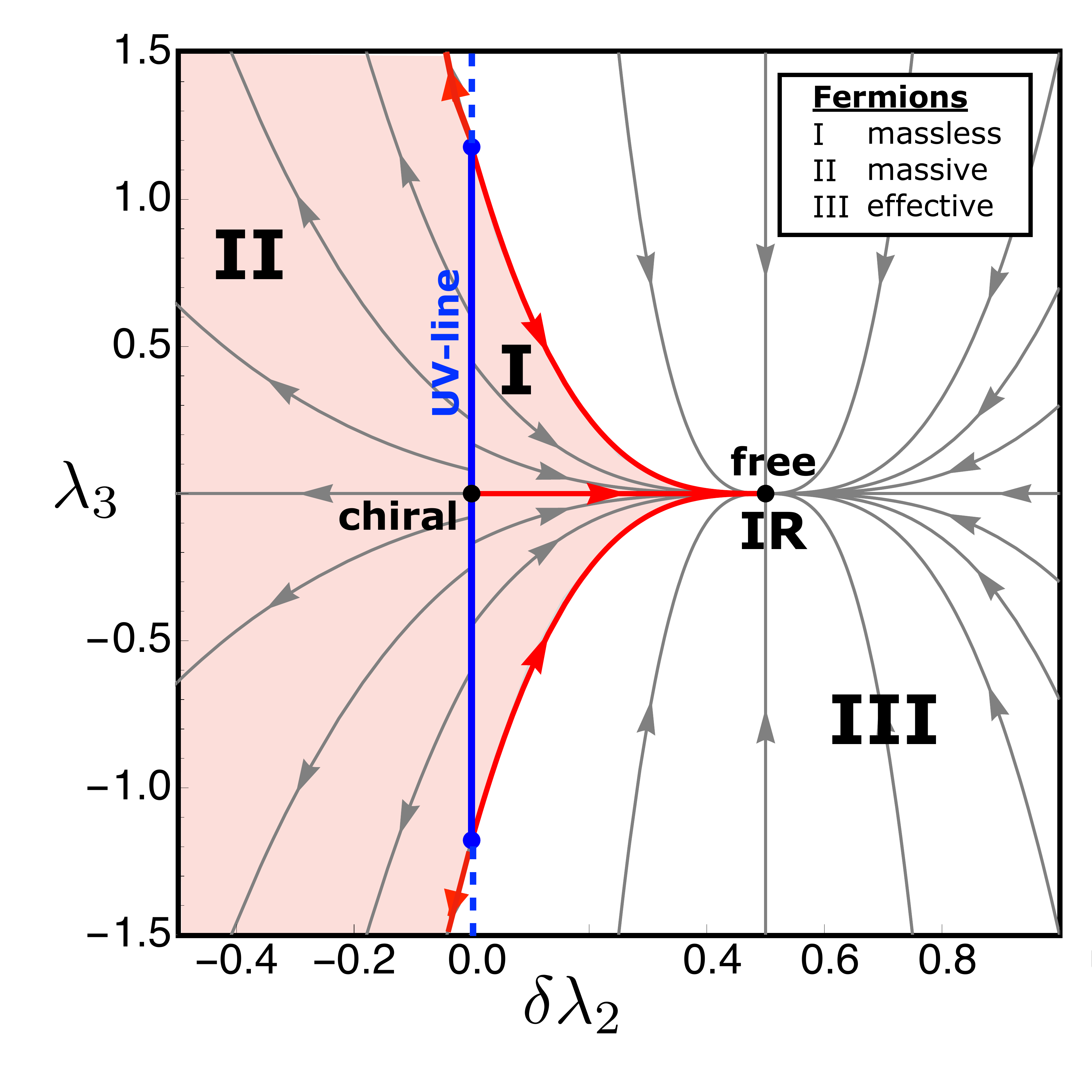}
\vskip-0.2cm
\caption{Phase diagram of the  fermionic theory \eq{eq:classical_action}
in the  $(\delta\la{2}, \la{3} )$ plane with $\delta\la{1}=0$. 
For $\la{3} \neq 0$ chiral symmetry is absent fundamentally, and ``emerges'' at the IR fixed point.}
\vskip-.4cm
\label{fig:l2_l3_flow}
\end{figure}

As soon as $\delta\la{1}= 0$, however, mass is not generated explicitly. This is noteworthy in that it shows that  chiral symmetry, which removes all $2n$F interactions $\la{n={\rm odd}}$,
is not necessary to  ensure massless fermions. Rather, the significantly milder constraint $\delta\la{1} = 0$ is already sufficient, courtesy of $\la{1}$ being natural, \eq{beta1}.
  
  Still, mass can   be generated dynamically through  strong interactions. 
  Parametrically, this
  takes the form of a second order quantum phase transition controlled by the microscopic parameter $\delta\la{2}$, with $\delta\la{2}<0$  ($\delta\la{2}>0$) leading to a massive (massless) phase.
In the presence of chiral symmetry \eqref{eq:discrete_symmetry},  strong 4F interactions are  responsible for the  generation of mass (Fig.~\ref{mass}, upper panel). 
In the absence of chiral symmetry,  6F interactions additionally enhance the mass with growing $|\la{3}|$, until it becomes indeterminate for finite sextic coupling   $|\la{3}|\to\la{3}^{\rm crit}$ right at the  endpoint of the conformal window (Fig.~\ref{mass} lower panel). 
We emphasize that mass is generated without the breaking of a discrete symmetry. Moreover, for  $\delta\la{2}>0$, we  observe the  ``emergence'' of chiral symmetry in the IR, curtesy of the fully attractive free  fixed point.
Lastly, mass can also be generated  explicitly  as soon as $\delta\la{1}\neq 0$, which then takes the form of a cross-over as a function of $\delta\la{2}$. 
\step

{\it Fermionic Phase Diagram.---}  We are now in a position to discuss the full phase diagram of the theory. Fig.~\ref{fig:l2_l3_flow} shows the UV-line of chirally non-symmetric fixed points  \eq{line} where $\la{3}^* \neq 0$, its endpoints, and the chirally symmetric UV and IR fixed points, all in the  $( \delta\la{2}, \la{3} )$-plane and for $\delta\la{1}=0$. Arrows 
point from the UV to the IR. 
Trajectories emanating  from the blue line correspond to UV-complete fundamental theories (red-shaded areas). 
In region I ($\delta\la{2}>0$ and $0 < |\la{3}^*| < \lambda_3^{\rm crit}$), they connect  interacting UV conformal fixed points  with the free theory in the IR. These theories remain strictly massless. 
Moreover, even though chiral symmetry is absent microscopically, it emerges in the IR.
In region II ($\delta\la{2}<0$ and $0 < |\la{3}^*| < \lambda_3^{\rm crit}$), 
 strong interactions lead to the dynamical generation of mass.
Here, chiral symmetry is manifestly absent at all scales, and 
trajectories connect UV  conformal fixed points with  massive Gross-Neveu theories in the IR.
Region III relates to all trajectories which do not start out at the UV-line. These ``swampland'' theories are not  UV-complete and must be seen as effective rather than fundamental.
As soon as  $\delta\la{1}\neq 0$, mass is also generated explicitly and  trajectories starting from the UV-line or in the swampland
invariably lead to massive  theories.
\step

\begin{figure}[t]
\includegraphics[width=.79\linewidth]{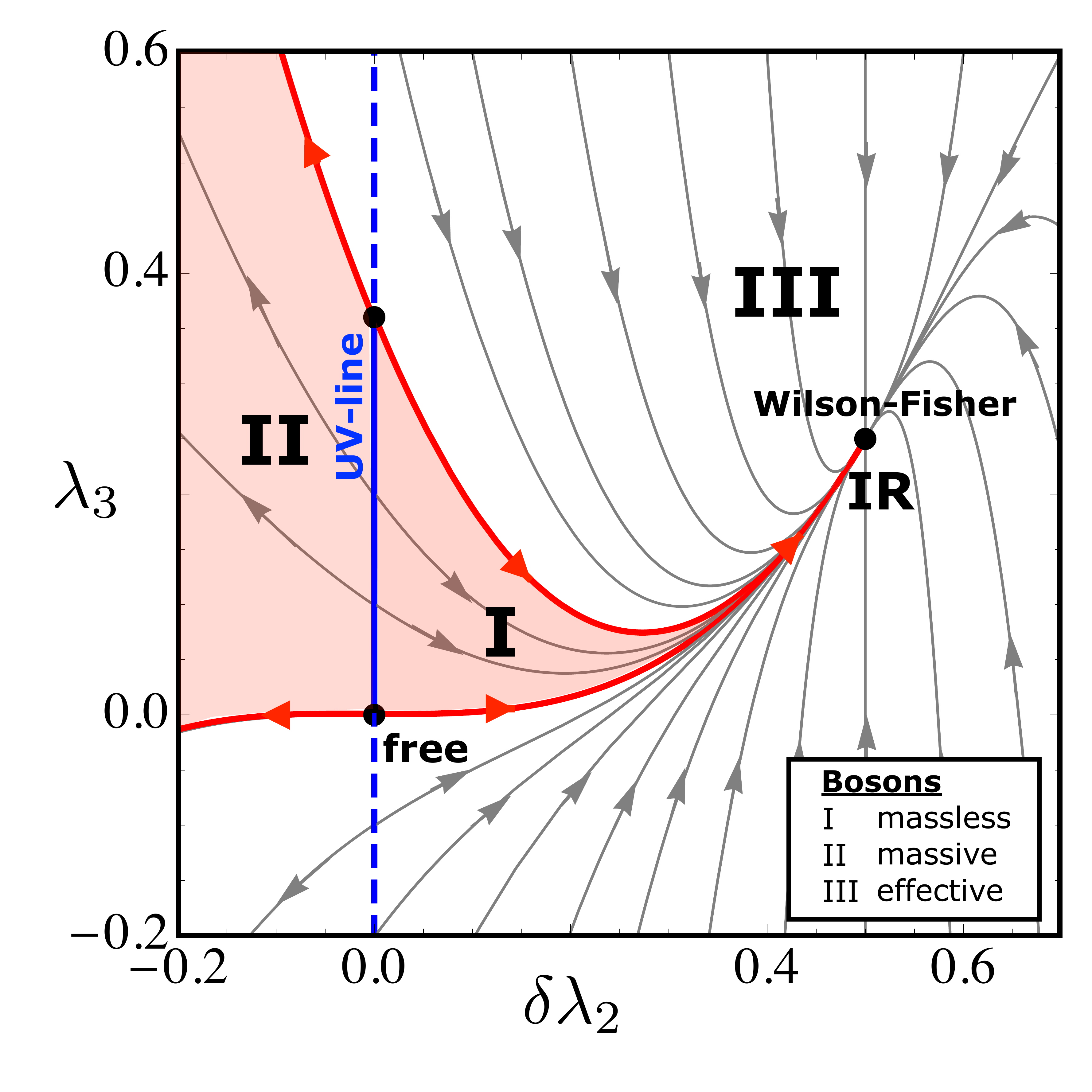}
\vskip-.1cm
\caption{Phase diagram of the  bosonic theory \eq{eq:classical_actionb} 
 in the  $(\delta\la{2}, \la{3} )$ plane with $\delta\la{1}=0$ (adapted from \cite{Litim:2017cnl}).}
\label{fig:bosonicPD}
\vskip-.2cm
\end{figure}

{\it Bosonic Phase Diagram.---} 
 It is interesting to compare our results 
 with  $O(N)$ [or $U(N)$] symmetric sextic scalar theory of 
 $3d$ real [or complex] bosons $\phi_a$ at large-$N$  \cite{David:1985zz,Litim:2017cnl,Litim:2018pxe}
with action
 \be\label{eq:classical_actionb}
S_{\rm b} = \! \!\int_x  
\Big\{ \frac12 ( \partial \phi)^2+\frac{M^2}2\phi^2+ \frac {G}2 (\phi^2)^2 + \frac{H}{3!}  (\phi^2)^3 \Big\}.
\ee
The bosonic theory is superrenormalisable  in the UV where it achieves a line of conformal fixed points  owing to an exactly marginal sextic scalar self-coupling  \cite{David:1985zz}. It also displays a strongly-interacting Wilson-Fisher fixed point in the IR (Fig.~\ref{fig:bosonicPD}).
In turn,  the fermionic theory \eq{eq:classical_action} is perturbatively non-renormalisable, yet it  develops a line of strongly interacting critical points in the UV owing to an exactly marginal six fermion coupling, alongside the isolated free  fixed point in the IR (Fig.~\ref{fig:l2_l3_flow}). 

Most notably,  we observe that the sets of UV and IR scaling dimensions 
of the bosonic theory \cite{Bardeen:1983st,David:1985zz,Litim:2017cnl,Litim:2018pxe} are {\it identical} to those of the  fermionic theory, \eq{IR} and \eq{UV}, respectively. 
We conclude that strongly interacting $3d$ fermions  in the UV can be viewed as ``quasi-bosons'' in that they display the same  conformal scaling  dimensions as free bosons along the entire  line of  fixed points. By the same token, strongly interacting  $3d$ scalars  can be viewed as ``quasi-fermions''. 

Comparing the  large $N$ phase diagrams and UV-IR connecting trajectories (Fig.~\ref{fig:l2_l3_flow} vs Fig.~\ref{fig:bosonicPD}), either of which has been obtained using  \eq{eq:wetterich}, we observe that both admit  massless (region I) and massive theories  (region II), and   theories without UV completion (region III).\footnote{In either of the phase diagrams, $\la{1,2,3}$ are the dimensonless counterparts of $M,G,H$  in \eq{eq:classical_action} and  \eq{eq:classical_actionb}, respectively,  and $\delta\lambda_2$  the deviation of $G$ from the critical point.}
Beyond large $N$, the line of fixed points 
collapse to a point. For the fermionic theory, this follows because the sextic becomes marginally irrelevant at $1/N$, which is mirrored in the scalar theory \cite{David:1985zz,Litim:2017cnl,Litim:2018pxe}. This is also in accord with  the fact that the Townsend-Pisarski fixed point for scalars  \cite{Townsend:1976sy,Pisarski:1982vz} and the  Gat-Kovner-Rosenstein fixed point  for fermions \cite{Gat:1991bf}
are both outside the stability domain 
at  order  $1/N$.
We take  these quantitative and structural similarities 
at large $N$ 
as a hint for a deeper 
link between the theories \eq{eq:classical_action} and \eq{eq:classical_actionb}
including away from critical points \cite{Cresswell-Hogg:2022lez}.
\step

{\it Discussion.---} 
Using the renormalisation group, we have  studied  large $N$ Gross-Neveu models  in three dimensions in the absence of chiral symmetry.
We find that the theory remains non-perturbatively renormalisable, much like its chirally-symmetric counterpart. The main novelty is  that  classically irrelevant 6F  interactions 
of mass dimension $-3$ have become exactly marginal due to quantum fluctuations. This is also in  agreement with earlier findings  \cite{Gat:1991bf} based on a Gross-Neveu-Yukawa version of the theory.
Hence, 
relaxing chiral symmetry 
adds  a new fundamentally free parameter, and opens
up an entire line of interacting UV fixed points. Fixed point solutions are well-defined globally, and limited by  endpoints where scale symmetry is broken spontaneously \cite{Cresswell-Hogg:2022lez}. 
Moreover, and even though fermion mass is no longer protected by symmetry, we find that massless theories 
prevail naturally, without any fine-tuning, 
and with chiral symmetry emerging in the IR. Further,
 the dynamical generation of mass, a combined effort of 4F and 6F interactions, 
 proceeds without the breaking of a discrete symmetry.

An intriguing aspect of our results are the striking  similarities with $3d$ scalar theories at large $N$. This includes equivalent conformal fixed points, phase diagrams, scaling dimensions, and UV-IR connecting trajectories. It will therefore be interesting to see whether  explicit maps can be found relating RG flows in the theory  of fermions \eq{eq:classical_action} to those in the theory of bosons \eq{eq:classical_actionb}, in the spirit of a large $N$ equivalence between a priori fundamentally different quantum field theories, e.g.~\cite{Bond:2019npq}. While many large $N$ equivalences or dualities in $3d$ involve Chern-Simons gauge fields, e.g.~\cite{Aharony:2012ns,Seiberg:2016gmd}, the latter make no appearance in our setup. In the spirit of \cite{Gat:1991bf}, it will  also be  interesting to cross-check  the findings
of this work within a Gross-Neveu-Yukawa  formulation of the theory   \cite{BMB},
given that   their critical points are expected to be equivalent.

Finally, we look at our findings from the viewpoint of conformal field theory or higher spin gauge theories. 
The  links between renormalisation group fixed points   and CFTs \cite{Cardy:1996xt}   
can be exploited to  extract   further conformal data  beyond scaling dimensions from our study. Moreover, certain versions of Vassiliev's higher spin theories on AdS${}_4$ have been shown to be dual to free or interacting large $N$ bosonic \cite{Klebanov:2002ja} or fermionic theories \cite{Sezgin:2003pt} on the boundary of AdS${}_4$, with the parity-even  Gross-Neveu UV fixed point ($\lambda_3=0)$ relating to  type-B Vassiliev theories \cite{Giombi:2012ms}. 
 Our study offers new large $N$ critical fermions without parity symmetry $(\lambda_3\neq 0)$. It would   also be interesting to understand whether critical endpoints  $(\lambda_3=\lambda_3^{\rm crit})$  continue to have higher spin duals \cite{Aharony:2012ns}, and whether the absence of parity or broken scale symmetry alters CFT three-point functions  of quasi-boson or quasi-fermion theories  \cite{Maldacena:2011jn,Maldacena:2012sf}. 
 \step

{\it Acknowledgements.---}  This work is supported by the Science and Technology Facilities Council (STFC) by a studentship (CCH), and under the Consolidated Grant ST/T00102X/1 (DFL).

\bibliography{6F}
\bibliographystyle{mystyle}

\end{document}